\documentclass[manuscript]{acmart}
\usepackage{hyperref}
\usepackage[T1]{fontenc}
\usepackage[utf8]{inputenc}
\usepackage{microtype}
\usepackage{natbib}
\usepackage{url}
\usepackage{amsthm}
\usepackage[inline]{enumitem}
\usepackage{multirow}
\usepackage{color, colortbl}
\usepackage{xcolor}
\usepackage{xspace}
\usepackage{soul}
\newcommand{\dataaugmethod}{\textsc{Mint}\xspace}

\copyrightyear{2023}
\acmYear{2023}
\setcopyright{acmlicensed}\acmConference[RecSys '23]{Seventeenth ACM Conference on Recommender Systems}{September 18--22, 2023}{Singapore, Singapore}
\acmBooktitle{Seventeenth ACM Conference on Recommender Systems (RecSys '23), September 18--22, 2023, Singapore, Singapore}
\acmPrice{15.00}
\acmDOI{10.1145/3604915.3608829}
\acmISBN{979-8-4007-0241-9/23/09}

\begin{document}

\title{Large Language Model Augmented Narrative Driven Recommendations}
\author{Sheshera Mysore}
\email{smysore@cs.umass.edu}
\affiliation{%
 \institution{University of Massachusetts Amherst}
 \country{USA}
}
\author{Andrew McCallum}
\email{mccallum@cs.umass.edu}
\affiliation{%
 \institution{University of Massachusetts Amherst}
 \country{USA}
}
\author{Hamed Zamani}
\email{hzamani@cs.umass.edu}
\affiliation{%
 \institution{University of Massachusetts Amherst}
 \country{USA}
}

\renewcommand{\shortauthors}{Mysore, McCallum, Zamani}
\begin{abstract}
Narrative-driven recommendation (NDR) presents an information access problem where users solicit recommendations with verbose descriptions of their preferences and context, for example, travelers soliciting recommendations for points of interest while describing their likes/dislikes and travel circumstances. These requests are increasingly important with the rise of natural language-based conversational interfaces for search and recommendation systems. However, NDR lacks abundant training data for models, and current platforms commonly do not support these requests. Fortunately, classical user-item interaction datasets contain rich textual data, e.g., reviews, which often describe user preferences and context -- this may be used to bootstrap training for NDR models. In this work, we explore using large language models (LLMs) for data augmentation to train NDR models. We use LLMs for authoring synthetic narrative queries from user-item interactions with few-shot prompting and train retrieval models for NDR on synthetic queries and user-item interaction data. Our experiments demonstrate that this is an effective strategy for training small-parameter retrieval models that outperform other retrieval and LLM baselines for narrative-driven recommendation.
\end{abstract}


\maketitle

\section{Introduction}
\label{sec-problem-goals}
\begin{figure}
    \centering
    \begin{minipage}{0.45\textwidth}
        \centering
        \includegraphics[width=\textwidth]{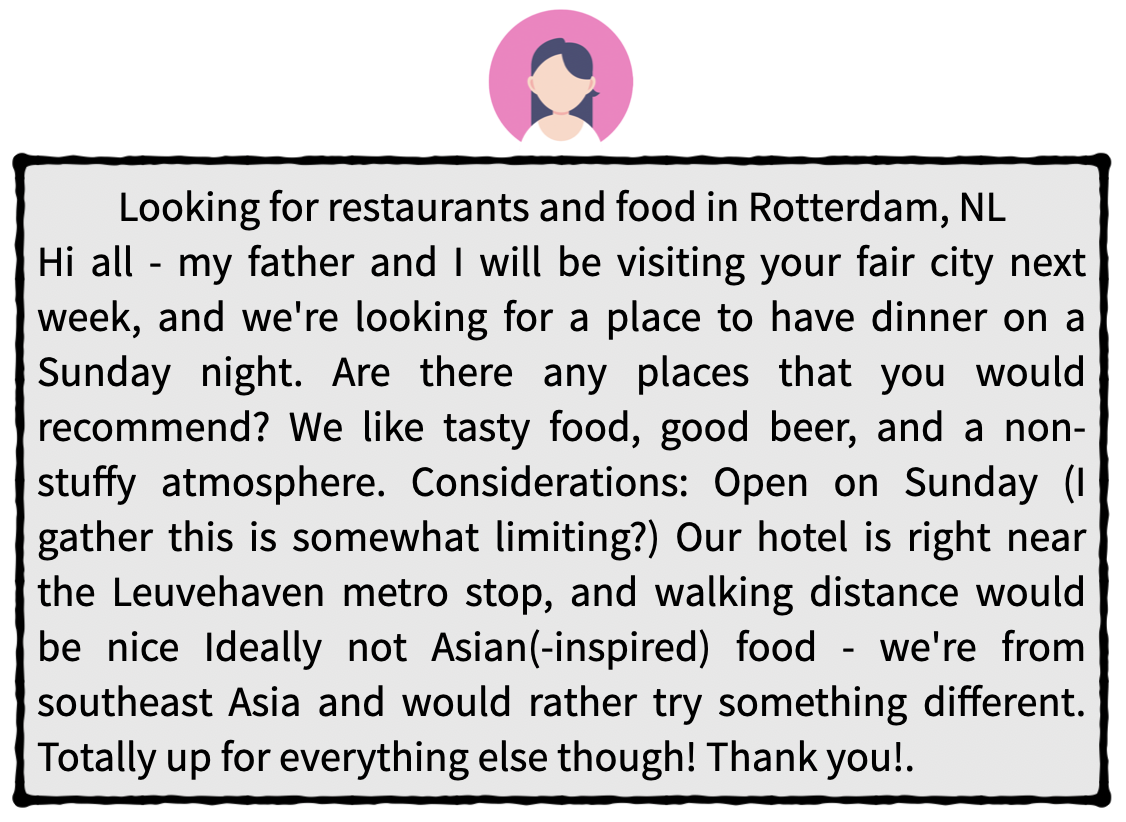} 
        \caption{An example narrative query soliciting point of interest recommendations. The query describes the users preferences and the context of their request.}
        \label{fig-ex-narrativeq}
        \Description{An example prompt used for generating narrative queries in MINT with GPT3. The prompt contains a few shot example with restaurant names and sentences from the reviews of the restaurants, and a hypothetical narrative query. An LLM is required to mimic the few shot example and generate a narrative query for a target set of items.}
    \end{minipage}\hfill
    \begin{minipage}{0.45\textwidth}
        \centering
        \includegraphics[width=\textwidth]{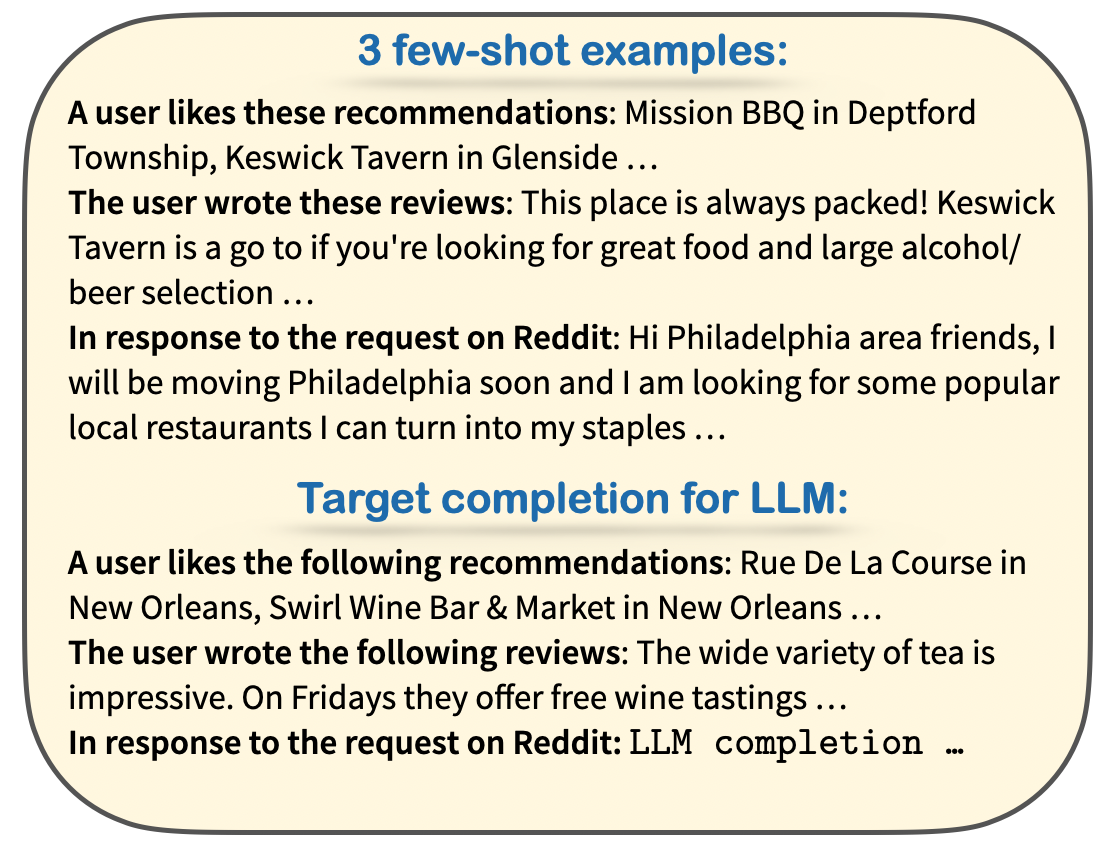} 
        \caption{The format of the prompt used in \dataaugmethod for generating synthetic narrative queries from user-item interaction with a large language model. }
        \label{fig-ex-promptformat}
    \end{minipage}
    \Description{An example narrative query of a user asking for restaurant recommendations in Rotterdam. The user describes an event, the kind of food they like, their cultural background and logistical restrictions.}
\end{figure}

Recommender systems personalized to users are an important component of several industry-scale platforms \cite{davidson2010youtube, das2007googlenes, jiajing2022pinterest}. These systems function by inferring users' interests from their prior interactions on the platform and making recommendations based on these inferred interests. While recommendations based on historical interactions are effective, users soliciting recommendations often start with a vague idea about their desired target items or may desire recommendations depending on the context of use, often missing in historical interaction data (Figure \ref{fig-ex-narrativeq}). In these scenarios, it is common for users to solicit recommendations through long-form narrative queries describing their broad interests and context. Information access tasks like these have been studied as narrative-driven recommendations (NDR) for items ranging from books \cite{bogers2017defining} and movies \cite{eberhard2019ndrreddit}, to points of interest \cite{afzali2021pointrec}. \citet{bogers2017defining} note these narrative requests to be common on discussion forums and several subreddits\footnote{\texttt{r/MovieSuggestions}, \texttt{r/booksuggestions}, \texttt{r/Animesuggest}}, but, there is a lack of support for these complex natural language queries in current recommenders. 

However, with the emergence of conversational interfaces for information access tasks, support for complex NDR tasks is likely to become necessary. In this context, recent work has noted an increase in complex and subjective natural language requests compared to more conventional search interfaces \cite{papenmeier2021convsearch, chen2022convnews}. Furthermore, the emergence of large language models (LLM) with strong language understanding capabilities presents the potential for fulfilling such complex requests \cite{brown2020gpt3, ouyang2022rlhf}. This work explores the potential for re-purposing historical user-item recommendation datasets, traditionally used for training collaborative filtering recommenders, with LLMs to support NDR.

Specifically, given a user's interactions, $D_u$, with items and their accompanying text documents (e.g., reviews, descriptions) $D_u=\{d_i\}_{i=1}^{N_u}$, selected from a user-item interaction dataset $\mathcal{I}$, we prompt \texttt{InstructGPT}, a 175B parameter LLM, to author a synthetic narrative query $q_u$ based on $D_u$ (Figure \ref{fig-ex-promptformat}). Since we expect the query $q_u$ to be noisy and not fully representative of all the user reviews, $D_u$ is filtered to retain only a fraction of the reviews based on a language-model assigned likelihood of $q_u$ given a user document, $d_i$. Then, a pre-trained LM based retrieval model (110M parameters) is fine-tuned for retrieval on the synthetic queries and filtered reviews. 

Our approach, which we refer to as \dataaugmethod\footnote{\dataaugmethod: Data aug\ul{M}entation with \ul{IN}teraction narra\ul{T}ives.}, follows from the observation that while narrative queries and suggestions are often made in online discussion forums, and could serve as training data, the number of these posts and the diversity of domains for which they are available is significantly smaller than the size and diversity of passively gathered user-item interaction datasets. E.g. while \citet{bogers2017defining} note nearly 25,000 narrative requests for books on the LibraryThing discussion forum, a publicly available user-item interaction dataset for Goodreads contains interactions with nearly 2.2M books by 460k users \cite{wan2018goodreads}
.

We empirically evaluate \dataaugmethod in a publicly available test collection for point of interest recommendation: \textsc{pointrec} \cite{afzali2021pointrec}. To train our NDR models, we generate synthetic training data based on user-item interaction datasets from Yelp. Models (110M parameters) trained with \dataaugmethod significantly outperform several baseline models and match the performance of significantly larger LLM baselines autoregressively generating recommendations. Code and synthetic datasets are available:\url{https://github.com/iesl/narrative-driven-rec-mint/}

\section{Related Work}
\label{sec-lit-review}
\textbf{Data Augmentation for Information Access.} A line of recent work has explored using language models to generate synthetic queries for data augmentation to train models for information retrieval tasks \cite{ma2021qgen, bonifacio2022inpars, dai2023promptagator, jeronymo2023inparsv2, boytsov2023inparslight}. Here, given a document collection of interest, a pre-trained language model is used to create synthetic queries for the document collection. An optional filtering step excludes noisy queries, and finally, a bi-encoder or a cross-encoder is trained for the retrieval task. While earlier work of \citet{ma2021qgen} train a custom query generation model on web-text datasets, more recent work has leveraged large language models for zero/few-shot question generation \cite{bonifacio2022inpars, dai2023promptagator, jeronymo2023inparsv2, boytsov2023inparslight}. In generating synthetic queries, this work indicates the effectiveness of smaller parameter LLMs (up to 6B parameters) for generating synthetic queries in simpler information-retrieval tasks  \cite{bonifacio2022inpars, jeronymo2023inparsv2, boytsov2023inparslight}, and finds larger models (100B parameters and above) to be necessary for harder tasks such as argument retrieval \cite{dai2023promptagator, jeronymo2023inparsv2}. 
Similar to this work, we explore the generation of synthetic queries with LLMs for a retrieval task. Unlike this work, we demonstrate a data augmentation method for creating effective training data from \emph{sets} of user documents found in recommendation datasets rather than individual documents. Other work in this space has also explored training more efficient multi-vector models from synthetic queries instead of more expensive cross-encoder models \cite{saadfalcon2023udapdr} and generating queries with a diverse range of intents than the ones available in implicit feedback datasets to enhance item retrievability \cite{penha2023retrievability}.

Besides creating queries for ad-hoc retrieval tasks, concurrent work of \citet{leszczynski2023generating} has also explored the creation of synthetic \emph{conversational} search datasets from music recommendation datasets with LLMs. The synthetic queries and user documents are then used to train bi-encoder retrieval models for conversational search. Our work resembles this in creating synthetic queries from \emph{sets} of user items found in recommendation interaction datasets. However, it differs in the task of focus, creating long-form narrative queries for NDR. Finally, our work also builds on the recent perspective of \citet{radlinski2022nlprofiles} who make a case for natural language user profiles driving recommenders -- narrative requests tie closely to natural language user profiles. Our work presents a step toward these systems.

Finally, while our work explores data augmentation from user-item interactions for a retrieval-oriented NDR task, prior work has also explored data augmentation of the user-item graph for training collaborative filtering models. This work has often explored augmentation to improve recommendation performance for minority \cite{ying2023camus, chen2023fairnessaug} or cold-start users \cite{wang2019genaug, chae2020arcf, lopez2021augmenting}. And has leveraged generative models \cite{wang2019genaug, chae2020arcf} and text similarity models \cite{lopez2021augmenting} for augmenting the user-item graph.

\textbf{Complex Queries in Information Access.} With the advent of performant models for text understanding, focus on complex and interactive information access tasks has seen a resurgence \cite{arguello2021tip, zamani2022conversational, lu2023statcan, mysore2021csfcube}. NDR presents an example of this -- NDR was first formalized in \citet{bogers2017defining} for the case of book recommendation and subsequently studied in other domains \cite{bogers2018m, bogers2019looking, bogers2018movie}. \citet{bogers2017defining} systematically examined narrative requests posted by users on discussion forums. They defined NDR as a task requiring item recommendation based on a long-form narrative query and prior-user item interactions. While this formulation resembles personalized search \cite{taveen2005personalizedsearch} and query-driven recommendation \cite{hariri2013qdriven}, the length and complexity of requests differentiate these from NDR. Other work has also demonstrated the effectiveness of re-ranking initial recommendations from collaborative filtering approaches based on the narrative query \cite{eberhard2019ndrreddit}. More recent work of \citet{afzali2021pointrec} formulate the NDR task without access to the prior interactions of a user while also noting the value of contextual cues contained in the narrative request. In our work, we focus on this latter formulation of NDR, given the lack of focus on effectively using the rich narrative queries in most prior work. Further, we demonstrate the usefulness of data augmentation from LLMs and user-item interaction datasets lacking narrative queries.

Besides this, a range of work has explored more complex, long-form, and interactive query formulations for information access; these resemble queries in NDR. \citet{arguello2021tip} define the tip of tongue retrieval task, a known-item search task where user queries describe the rich context of items while being unable to recall item metadata itself. \citet{mysore2021csfcube} formulate an aspect conditional query-by example task where results must match specific aspects of a long natural language query. And finally, a vibrant body of work has explored conversational critiquing of recommenders where natural language feedback helps tune the recommendations received by users \cite{zou2020questiondriven, wang2021controllable, luo2020latentlinear}. 

\section{Method}
\label{sec-method}

\subsection{Problem Setup}
\label{sec-problem-formulation}
In our work, we define narrative-driven recommendation (NDR) to be a ranking task, where given a narrative query $q$ made by a user $u$, a ranking system $f$ must generate a ranking $R$ over a collection of items $\mathcal{C}$. Further, we assume access to a user-item interaction dataset $\mathcal{I}$ consisting of user interactions with items $(u, \{d_i\}_{i=1}^{N_u})$. We assume the items $d_i$ to be textual documents like reviews or item descriptions. While we don't assume there to be any overlap in the users making narrative queries or the collection of items $\mathcal{C}$ and the user-items interaction dataset $\mathcal{I}$, we assume them to be from the same broad domain, e.g., books, movies, points-of-interest.

\subsection{Proposed Method}
\label{sec-proposed-method}
\begin{figure}[t]
     \centering
     \fbox{\includegraphics[width=0.9\textwidth]{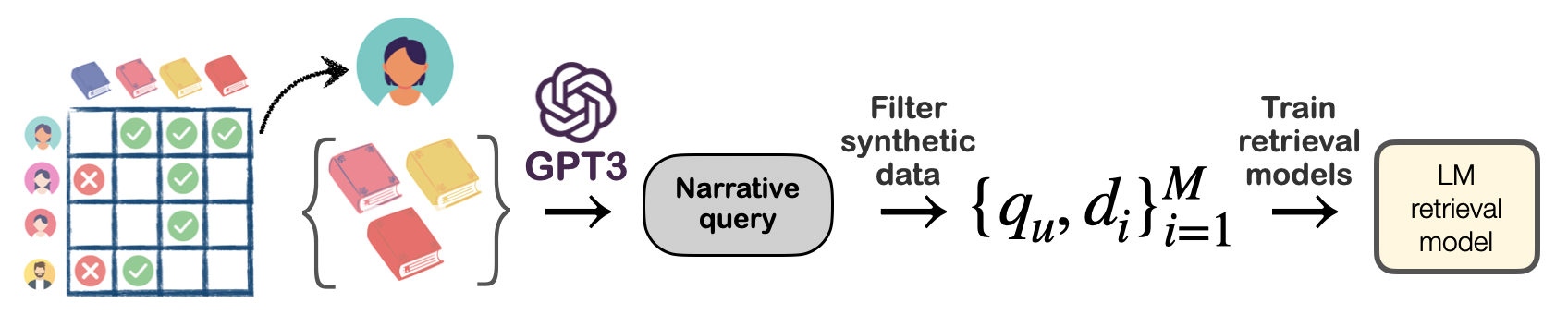}}
     \caption{\dataaugmethod re-purposes readily available user-item interaction datasets commonly used to train collaborative filtering models for narrative-driven recommendation. This is done by authoring narrative queries for sets of items liked by a user with a large language model. The data is filtered with a smaller language model and retrieval models are trained on the synthetic queries and user items.}
     \label{fig-high-level}
     \Description{A schematic for the proposed method. The figure depicts how user-items may be selected from a interaction matric, used for authoring queries with GPT-3, and then used to train a retrieval model for narrative driven recommendation.}
 \end{figure}

Our proposed method, \dataaugmethod, for NDR, re-purposes a dataset of abundantly available user-item interactions, $\mathcal{I} = \{(u, \{d_i\}_{i=1}^{N_u})\}$ into training data for retrieval models by using LLMs as query generation models to author narrative queries $q_u$: $\mathcal{D} = \{(q_u, \{d_i\}_{i=1}^{N_u})\}$. Then, retrieval models are trained on the synthetic dataset $\mathcal{D}$ (Figure \ref{fig-high-level}).

\subsubsection{Narrative Queries from LLMs}
To author a narrative query $q_u$ for a user in $\mathcal{I}$, we make use of the 175B parameter \texttt{InstructGPT}\footnote{\url{https://platform.openai.com/docs/models/gpt-3}, \texttt{text-davinci-003}} model as our query generation model $\textsc{QGen}$. We include the text of interacted items $\{d_i\}_{i=1}^{N_u}$ in the prompt for $\textsc{QGen}$, and instruct it to author a narrative query (Figure \ref{fig-ex-promptformat}). To improve the coherence of generated queries and obtain correctly formatted outputs, we manually author  narrative queries for 3 topically diverse users based on their interacted items and include it in the prompt for $\textsc{QGen}$. The same three few shot examples are used for the whole dataset $\mathcal{I}$, and the three users were chosen from $\mathcal{I}$. 
Generating narrative queries based on user interactions may also be considered a form of multi-document summarization for generating a natural language user profile \cite{radlinski2022nlprofiles}.

\subsubsection{Filtering Items for Synthetic Queries} 
\label{sec-filtering-items}
Since we expect user items to capture multiple aspects of their interests and generated queries to only capture a subset of these interests, we only retain some of the items present in $\{d_i\}_{i=1}^{N_u}$ before using it for training retrieval models. For this, we use a pre-trained language model to compute the likelihood of the query given each user item, $P_{LM}(q_u|d_i)$, and only retain the top $M$ highly scoring item for $q_u$, this results in $M$ training samples per user for our NDR retrieval models: $\{(q_u, d_i)_{i=1}^{M}\}$. In our experiments, we use \textsc{FlanT5} with 3B parameters \cite{chung2022scaling} for computing and follow \citet{sachan2022upr} for computing $P_{LM}(q_u|d_i)$. Note that our use of $P_{LM}(q_u|d_i)$ represents a query-likelihood model classically used for ad-hoc search and recently shown to be an effective unsupervised re-ranking method when used with large pre-trained language models \cite{sachan2022upr}.

\subsubsection{Training Retrieval Models} 
\label{sec-training}
We train bi-encoder and cross-encoder models for NDR on the generated synthetic dataset -- commonly used models in search tasks. Bi-encoders are commonly used as scalable first-stage rankers from a large collection of items. On the other hand, cross-encoders allow a richer interaction between query and item and are used as second-stage re-ranking models. For both models, we use a pre-trained transformer language model architecture with 110M parameters, \textsc{MPnet}, a model similar to \textsc{Bert} \cite{song2020mpnet}. Bi-encoder models embed the query and item independently into high dimensional vectors: $\mathbf{q}_u=\textsc{MPNet}(q_u)$, $\mathbf{d}_i=\textsc{MPNet}(d_i)$ and rank items for the user based on the minimum L2 distance between $\mathbf{q}_u$ and $\mathbf{d}_i$. Embeddings are obtained by averaging token embeddings from the final layer of $\textsc{MPNet}$, and the same model is used for both queries and items. Cross-encoder models input both the query and item and output a score to be used for ranking $s=f_{\textsc{Cr}}([q_u; d_i])$, where $f_{\textsc{Cr}}$ is parameterized as $\mathbf{w}^T\texttt{dropout}\left(\mathbf{W}^T\textsc{MPNet}(\cdot)\right)$. We train our bi-encoder model with a margin ranking loss: $\mathcal{L}_{Bi} = \sum_{u}\sum_{i=1}^{M}\texttt{max}[L2(\textbf{q}_u, \textbf{d}_i)-L2(\textbf{q}_u, \textbf{d}^{'}_{i})+\delta,0]$ with randomly sampled negatives $d^{'}$ and $\delta=1$. Our cross-encoders are trained with a cross-entropy loss: $\mathcal{L}_{Cr}=\sum_{u}\sum_{i=1}^{M}\texttt{log}(\frac{e^s}{\sum_{d'}e^{s'}})$. For training, 4 negative example items $d'$ are randomly sampled from ranks 100-300 from our trained bi-encoder. At test time, we retrieve the top 200 items with our trained bi-encoder and re-rank them with the cross-encoder - we evaluate both these components in experiments and refer to them as \texttt{BiEnc}-\dataaugmethod and \texttt{CrEnc}-\dataaugmethod.

\section{Experiments and Results}
\label{sec-results}
Next, we evaluate \dataaugmethod on a publicly available test collection for NDR and present a series of ablations.

\subsection{Experimental Setup}
\label{sec-exp-setup}

\subsubsection{Datasets}
\label{sec-results}
We perform evaluations on an NDR dataset for point-of-interest (POI) recommendation \textsc{Pointrec} \cite{afzali2021pointrec}. \textsc{Pointrec} contains 112 realistic narrative queries (130 words long) obtained from discussion forums on Reddit and items pooled from baseline rankers. The items are annotated on a graded relevance scale by crowd-workers and/or discussion forum members and further validated by the dataset authors. The item collection $\mathcal{C}$ in \textsc{Pointrec} contains 700k POIs with metadata (category, city) and noisy text snippets describing the POI obtained from the Bing search engine. For test time ranking, we only rank the candidate items in the city and request category (e.g., ``Restaurants'') of the query available in \textsc{Pointrec} - this follows prior practice to exclude clearly irrelevant items \cite{liu2013preftrans, afzali2021pointrec}. 
We use user-item interaction datasets from Yelp to generate synthetic queries for training.\footnote{\url{https://www.yelp.com/dataset}}
Note also that we limit our evaluations to \textsc{Pointrec} since it presents the only publicly available, manually annotated, and candidate pooled test collection for NDR, to our knowledge. Other datasets for NDR use document collections that are no longer publicly accessible \cite{koolen2016clefsvs}, contain sparse and noisy relevance judgments due to them being determined with automatic rules applied to discussion threads \cite{eberhard2019ndrreddit, koolen2016clefsvs}, lack pooling to gather candidates for judging relevance \cite{eberhard2019ndrreddit, koolen2016clefsvs}, or lack realistic narrative queries \cite{hashemi2016overview}. We leave the development of more robust test collections and evaluation methods for NDR to future work.

\subsubsection{Implementation Details}
\label{sec-impl-details}
Next, we describe important details for \dataaugmethod and leave finer details of the model and training to our code release. To sample user interactions for generating synthetic queries from the Yelp dataset, we exclude POIs and users with fewer than ten reviews to ensure that users were regular users of the site with well represented interests. This follows common prior practice in preparing user-item interaction datasets for use \cite{liu2017poiprec}. Then we retain users who deliver an average rating greater than $3/5$ and with 10-30 above-average reviews. This desirably biases our data to users who commonly describe their likings (rather than dislikes). It also retains the users whose interests are summarizable by \textsc{QGen}. In the Yelp dataset, this results in 45,193 retained users. Now, 10,000 randomly selected users are chosen for generating synthetic narrative queries. For these users, a single randomly selected sentence from 10 of their reviews is included in the prompt (Figure \ref{fig-ex-promptformat}) to \textsc{QGen}, i.e., \ $N_u=10$. After generating synthetic queries, some items are filtered out (\S\ref{sec-filtering-items}). Here, we exclude 40\% of the items for a user. This results in about 60,000 training samples for training \texttt{BiEnc}-\dataaugmethod and \texttt{CrEnc}-\dataaugmethod. These decisions were made manually by examining the resulting datasets and the cost of authoring queries. The expense of generating $q_u$ was about USD 230.

\subsubsection{Baselines}
\label{sec-baselines}
We compare \texttt{BiEnc}-\dataaugmethod and \texttt{CrEnc}-\dataaugmethod models against several standard and performant retrieval model baselines. These span zero-shot/unsupervised rankers, supervised bi-encoders, unsupervised cross-encoders, and LLM baselines. \ul{BM25}: A standard unsupervised sparse retrieval baseline based on term overlap between query and document, with strong generalization performance across tasks and domains \cite{robertson2009bm25}. \ul{Contriver}: A BERT-base bi-encoder model pre-trained for zero-shot retrieval with weakly supervised query-document pairs \cite{izacard2022unsupervised}. \ul{MPNet-1B}: A strong Sentence-Bert bi-encoder model initialized with MPNet-base and trained on 1 billion supervised query-document pairs aggregated from numerous domains \cite{reimers2019sentencebert}. \ul{BERT-MSM}: A BERT-base bi-encoder fine-tuned on supervised question-passage pairs from \textsc{MSMarco}. \ul{UPR}: A two-stage approach that retrieves items with a Contriver bi-encoder and re-ranks the top 200 items with a query-likelihood model using a FlanT5 model with 3B parameters \cite{sachan2022upr, chung2022scaling}. This may be seen as an unsupervised ``cross-encoder'' model. \ul{Grounded LLM}: A recently proposed two-stage approach which autoregressively generates ten pseudo-relevant items using an LLM (175B \textsc{InstructGPT}) prompted with the narrative query and generates recommendations grounded in $\mathcal{C}$ by retrieving the nearest neighbors for each generated item using a bi-encoder \cite{gao2022precise}. We include one few-shot example of a narrative query and recommended items in the prompt to the LLM. We run this baseline three times and report average performance across runs. We report NDCG at 5 and 10, MAP, MRR, and Recall at 100 and 200. Finally, our reported results should be considered lower bounds on realistic performance due to the unjudged documents (about 70\% at $k=10$) in our test collections \cite{buckley2004incomplete}.
\begin{table}[]
\caption{Performance of the proposed method, \dataaugmethod, for point-of-interest recommendation on \textsc{Pointrec}. The superscripts denote statistically significant improvements compared to specific baseline models.}
\scalebox{1}{\begin{tabular}{rcllllll}
& & \multicolumn{6}{c}{\textsc{Pointrec}}\\
\cmidrule(lr){3-8} 
 Model &  Parameters  & \small{NDCG@5}   & \small{NDCG@10} & \small{MAP}   & \small{MRR} & \small{Recall@100} & \small{Recall@200}\\\toprule
$^1$BM25 & - & 0.2682 & 0.2464 & 0.1182 & 0.2685 & 0.4194 & 0.5429\\
$^2$Contriver & 110M & 0.2924 & 0.2776 & 0.1660 & 0.3355 & 0.4455 & 0.5552\\
$^3$MPNet-1B  & 110M & 0.3038 & 0.2842 & 0.1621 & 0.3566 & 0.4439 & 0.5657\\
$^4$BERT-MSM  & 110M & 0.3117 & 0.2886 & 0.1528 & 0.3320 & 0.4679 & 0.5816\\\midrule
$^5$Grounded LLM & 175B$+$110M & 0.3558 & 0.3251 & 0.1808 & 0.3861 & 0.4797 & 0.5797\\
$^6$UPR & 110M$+$3B & 0.3586 & 0.3242 & 0.1712 & 0.4013 & 0.4489 & 0.5552\\\midrule
\texttt{BiEnc}-\dataaugmethod  & 110M & 0.3489$^1$ & 0.3263$^1$ & 0.1890$^1$ & 0.3982$^1$ & 0.4914$^1$ & \textbf{0.6221}\\
\texttt{CrEnc}-\dataaugmethod & 2$\times$110M & \textbf{0.3725}$^{12}$ & \textbf{0.3489}$^{12}$ & \textbf{0.2192}$^{14}$ & \textbf{0.4317}$^{1}$ & \textbf{0.5448}$^{123}$ & \textbf{0.6221}\\
\bottomrule 
\end{tabular}}
\label{table-ndr-main}
\Description[Main results for MINT]{This table presents evaluations for bi-encoders and crossencoders trained on Mint data on the point-of-interest dataset Pointrec and compared to a host of baselines and their parameter counts.}
\end{table}

\subsection{Results}
\label{sec-results}
Table \ref{table-ndr-main} presents the performance of the proposed method compared against baselines. Here, bold numbers indicate the best-performing model, and superscripts indicate statistical significance computed with two-sided t-tests at $p<0.05$.

Here, we first note the performance of baseline approaches. We see BM25 outperformed by Contriver, a transformer bi-encoder model trained for zero-shot retrieval; this mirrors prior work \cite{izacard2022unsupervised}. Next, we see supervised bi-encoder models trained on similar passage (MPNet-1B) and question-answer (BERT-MSM) pairs outperform a weakly supervised model (Contriver) by smaller margins. Finally, the Grounded LLM outperforms all bi-encoder baselines, indicating strong few-shot generalization and mirroring prior results \cite{gao2022precise}. Examining the \dataaugmethod models, we first note that the \texttt{BiEnc}-\dataaugmethod sees statistically significant improvement compared to BM25 and outperforms the best bi-encoder baselines by 11-13\% on precision measures and 5-7\% on recall measures. Specifically, we see a model trained for question-answering (BERT-MSM) underperform \texttt{BiEnc}-\dataaugmethod, indicating the challenge of the NDR task. Further, \texttt{BiEnc}-\dataaugmethod, trained on 5 orders of magnitude lesser data than \textsc{MPNet-1B}, sees improved performance -- indicating the quality of data obtained from \dataaugmethod. Furthermore, \texttt{BiEnc}-\dataaugmethod also performs at par with a 175B LLM while offering the inference efficiency of a small-parameter bi-encoder. Next, we see \texttt{CrEnc}-\dataaugmethod outperform the baseline bi-encoders, \texttt{BiEnc}-\dataaugmethod, UPR, and Grounded LLM by 4-21\% on precision measures and 7-13\% on recall measures -- demonstrating the value of augmentation with \dataaugmethod for training NDR models.

\subsection{Ablations}
\label{sec-ablations}
In Table \ref{table-ndr-ablation}, we ablate various design choices in \dataaugmethod. Different choices result in different training sets for the \texttt{BiEnc} and \texttt{CrEnc} models. Also, note that in reporting ablation performance for \texttt{CrEnc}, we still use the performant \texttt{BiEnc}-\dataaugmethod model for obtaining negative examples for training and first-stage ranking. Without high-quality negative examples, we found \texttt{CrEnc} to result in much poorer performance.

\ul{No item filtering.} Since synthetic queries are unlikely to represent all the items of a user, \dataaugmethod excludes user items $\{d_i\}_{i=1}^{N_u}$ which have a low likelihood of being generated from the document (\S\ref{sec-filtering-items}). Without this step, we expect the training set for training retrieval models to be larger and noisier. In Table \ref{table-ndr-ablation}, we see that excluding this step leads to a lower performance for \texttt{BiEnc} and \texttt{CrEnc}, indicating that the quality of data obtained is important for performance.

\ul{6B LLM for \textsc{QGen}.} \dataaugmethod relies on using an expensive 175B parameter \textsc{InstructGPT} model for \textsc{QGen}. Here, we investigate the efficacy for generating $q_u$ for $\{d_i\}_{i=1}^{N_u}$ with a 6B parameter \textsc{InstructGPT} model (\texttt{text-curie-001}). We use an identical setup to the 175B LLM for this. In Table \ref{table-ndr-ablation}, we see that training on the synthetic narrative queries of the smaller LLM results in worse models -- often underperforming the baselines in Table \ref{table-ndr-main}. This indicates the inability of a smaller model to generate complex narrative queries while conditioning on a set of user items. This necessity of a larger LLM for generating queries in complex retrieval tasks has been observed in prior work \cite{dai2023promptagator, jeronymo2023inparsv2}.

\ul{6B LLM for Item Queries.} We find a smaller 6B LLM to result in poor quality data when used to generate narrative queries conditioned on $\{d_i\}_{i=1}^{N_u}$. Here we simplify the text generation task -- using a 6B LLM to generate queries for individual items $d_i$. This experiment also mirrors the setup for generating synthetic queries for search tasks \cite{dai2023promptagator, bonifacio2022inpars}. Here, we use 3-few shot examples and sample one item per user for generating $q_u$. Given the lower cost of using a smaller LLM, we use all 45,193 users in our Yelp dataset rather than a smaller random sample. From Table \ref{table-ndr-ablation}, we see that this results in higher quality queries than using smaller LLMs for generating narrative queries from $\{d_i\}_{i=1}^{N_u}$. The resulting \texttt{BiEnc} model underperforms the \texttt{BiEnc}-\dataaugmethod, indicating the value of generating complex queries conditioned on multiple items as in \dataaugmethod for NDR. We see that \texttt{CrEnc} approaches the performance of \texttt{CrEnc}-\dataaugmethod -- note, however, that this approach uses the performant \texttt{BiEnc}-\dataaugmethod for sampling negatives and first stage ranking. We leave further exploration of using small parameter LLMs for data augmentation for NDR models to future work.
\begin{table}[]
\caption{\dataaugmethod ablated for different design choices on \textsc{Pointrec}.}
\scalebox{1}{\begin{tabular}{lllllll}
& \multicolumn{6}{c}{\textsc{Pointrec}}\\
\cmidrule(lr){2-7}
 Ablation & \small{NDCG@5}   & \small{NDCG@10} & \small{MAP}   & \small{MRR} & \small{Recall@100} & \small{Recall@200}\\\toprule
\texttt{BiEnc}-\dataaugmethod & 0.3489 & 0.3263 & 0.1890 & 0.3982 & 0.5263 & 0.6221\\
$-$ No item filtering & 0.2949 & 0.2766 & 0.1634 & 0.3505 & 0.4979 & 0.5951\\
$-$ 6B LLM for $\textsc{QGen}$ & 0.2336 & 0.2293 & 0.1125 & 0.2287 & 0.426 & 0.5435\\
$-$ 6B LLM for Item Queries & 0.3012 & 0.2875 & 0.1721 & 0.3384 & 0.4800 & 0.5909\\\midrule
\texttt{CrEnc}-\dataaugmethod & {0.3725} & {0.3489} & {0.2192} & {0.4317} & {0.5448} & {0.6221}\\
$-$ No item filtering & 0.3570 & 0.3379 & 0.2071 & 0.4063 & 0.5366 & 0.6221\\
$-$ 6B LLM for $\textsc{QGen}$ & 0.2618 & 0.2421 & 0.1341 & 0.3118 & 0.4841 & 0.6221\\
$-$ 6B LLM for Item Queries & 0.3792 & 0.3451 & 0.2128 & 0.4098 & 0.5546 & 0.6221\\
\bottomrule 
\end{tabular}}
\label{table-ndr-ablation}
\Description[Ablation table]{Ablations comparing Mint bi-encoders and cross-encoders to models trained on queries from smaller large language models, item queries, and unfiltered items in a users profile.}
\end{table}

\section{Conclusions}
\label{sec-conclusion}
In this paper, we present \dataaugmethod, a data augmentation method for the narrative-driven recommendation (NDR) task. \dataaugmethod re-purposes historical user-item interaction datasets for NDR by using a 175B parameter large language model to author long-form narrative queries while conditioning on the text of items liked by users. We evaluate bi-encoder and cross-encoder models trained on data from \dataaugmethod on the publicly available \textsc{Pointrec} test collection for narrative-driven point of interest recommendation. We demonstrate that the resulting models outperform several strong baselines and ablated models and match or outperform a 175B LLM directly used for NDR in a 1-shot setup.

However, \dataaugmethod also presents some limitations. Given our use of historical interaction datasets for generating synthetic training data and the prevalence of popular interests in these datasets longer, tailed interests are unlikely to be present in the generated synthetic datasets. In turn, causing retrieval models to likely see poorer performance on these requests. Our use of LLMs to generate synthetic queries also causes the queries to be repetitive in structure, likely causing novel longer-tail queries to be poorly served. These limitations may be addressed in future work. 

Besides this, other avenues also present rich future work. While \dataaugmethod leverages a 175B LLM for generating synthetic queries, smaller parameter LLMs may be explored for this purpose - perhaps by training dedicated \textsc{QGen} models. \dataaugmethod may also be expanded to explore more active strategies for sampling items and users for whom narrative queries are authored - this may allow more efficient use of large parameter LLMs while ensuring higher quality training datasets. Next, the generation of synthetic queries from \textit{sets} of documents may be explored for a broader range of retrieval tasks beyond NDR given its promise to generate larger training sets -- a currently underexplored direction. Finally, given the lack of larger-scale test collections for NDR and the effectiveness of LLMs for authoring narrative queries from user-item interaction, fruitful future work may also explore the creation of larger-scale datasets in a mixed-initiative setup to robustly evaluate models for NDR tasks.

\begin{acks}
We thank anonymous reviewers for their invaluable feedback. This work was partly supported by the Center for Intelligent Information Retrieval, NSF grants IIS-1922090 and 2143434, the Office of Naval Research contract number N000142212688, an Amazon Alexa Prize grant, and the Chan Zuckerberg Initiative under the project Scientific Knowledge Base Construction. Any opinions, findings and conclusions or recommendations expressed in this material are those of the authors and do not necessarily reflect those of the sponsors.    
\end{acks}


\bibliographystyle{ACM-Reference-Format}
\bibliography{sample-base}

\end{document}